\documentclass[12pt,twoside]{article}
\usepackage[dvips]{color}
\usepackage{graphicx}
\usepackage[figuresright]{rotating}

\newcommand{\beq}{\begin{equation}}
\newcommand{\eeq}{\end{equation}}
\newcommand{\beqn}{\begin{eqnarray}}
\newcommand{\eeqn}{\end{eqnarray}}  
\definecolor{darkmagenta}{rgb}{1,0,.5}
\begin{document}
\title{Search for T-Violation in 
 $K^+\rightarrow\pi^0\mu^+\nu$ and $K^+\rightarrow\mu^+\nu\gamma$ 
Decays }

\author{Yu.G.~Kudenko \\
Institute for Nuclear Research RAS, 117312 Moscow, Russia  \\}	 
%\begin{document}
\date{}
\maketitle
\begin{abstract}
The recent progress in
a search for T-violating transverse muon polarization in the decays 
$K^{+}\rightarrow\pi^{0}\mu^{+}\nu$ and 
$K^+\rightarrow\mu^+\nu\gamma$ in
the on-going experiment E246 at KEK is reported. Future 
prospects  
in polarization measurements are also discussed. 
\end{abstract} 
\section{Introduction}

Measurement of the muon transverse polarization, $P_T$,    
 in the decays
$K^+\to\pi^0\mu^+\nu$~($K_{\mu3}$) and  
$K^+\to\mu^+\nu\gamma$~($K_{\mu2\gamma}$) can provide a clear signature  
 of new physics beyond the Standard Model (SM).  
 In  
$K_{\mu3}$ decay, $P_T$ is a T-odd observable ${\bf s}_{\mu}\cdot
({\bf p}_{\pi}\times{\bf p}_{\mu})$ determined by the
$\pi^0$ momentum ${\bf p}_{\pi}$ and the muon  
momentum  
${\bf p}_{\mu}$ and spin ${\bf s}_{\mu}$.  
In the case of
$K_{\mu2\gamma}$ decay, $P_T$ is proportional to ${\bf s}_{\mu}\cdot
({\bf q}\times{\bf p}_{\mu})$, where ${\bf q}$ is the 
photon momentum.  
 These observables are very small in the SM~\cite{valencia}, but
they are interesting probes of various non-SM CP-violation
mechanisms~\cite{nonsm,geng,kob} where $P_T$  
could  
 be as large as 
$10^{-3}$
in either $K_{\mu3}$ or $K_{\mu2\gamma}$.
However,
the  electromagnetic final-state
interactions (FSI)  can generate  a physics background, 
i.e. nonvanishing $P_T$, in both decays. 
The value of $P_T$
due to the FSI is expected to be about  
$4\times 10^{-6}$ for    
$K_{\mu3}$ decay~\cite{zhi,kmu3fsi},
i.e.,  
 much smaller than the expected non-SM effects. On the other hand, 
for the $K_{\mu2\gamma}$ decay, the FSI can
induce  a $P_T$ of  $\leq 10^{-3}$~\cite{efrkud}. In this decay, 
$P_T$ due to FSI depends on the 
values of 
an axial vector form factor $F_A$  and a vector form factor $F_V$.  
Moreover, $P_T$ varies significantly over the Dalitz plot,  
 reaching a  
   maximum value of 
$10^{-3}-10^{-2}$ 
at large muon energy, i.e.,  
  in the region of high sensitivity to 
T-violating
parameters~\cite{geng}.

For  $K_{\mu3}$ decay,  
the most general invariant amplitude is
\beqn
\label{kmu3ampl/1} 
M_{K_{\mu3}} & = & -\frac{G_F}{\sqrt2}V_{us}
 \frac{1}{\sqrt2}\Bigl[(p_K +p_{\pi} )_{\lambda} f_+(t)+(p_K
-p_{\pi})_{\lambda} f_-(t) \Bigr]  \nonumber\\
& & \times \bar{u}(p_{\mu}) \gamma_{\lambda}
(1+\gamma_5 )v(p_{\nu}).
\eeqn
Here $G_F$ is the Fermi constant; $V_{us}$ is the 
Kobayashi-Maskawa matrix element;  
$p_K,\; p_{\pi},\; p_{\mu},$ and
$p_{\nu}$ are the momenta of the kaon, pion, muon, and
antineutrino, respectively; $t=(p_K -p_{\pi})^2$ is the momentum
transfer to the lepton pair squared.
In the $K^+$ center of mass system,  $P_{T}$ can be expressed as a 
function
of the muon and pion energies
\beq
P_{T} \cong 
Im(\xi)(\frac{m_{\mu}}{m_{K}})\frac{|\vec{p_{\mu}}|}{[E_{\mu} + 
|\vec{p_{\mu}}|\vec{n_{\mu}}\cdot\vec{n_{\nu}} - m^{2}_{\mu}/m_{K}]} 
=Im(\xi)\cdot\Phi,
\eeq 
where the parameter $\xi(q^2) = f_{-}(q^2)/f_{+}(q^2)$, $m_{\mu}$ 
and  $m_{K}$ are the muon and kaon mass, respectively,   
 $\vec{n_{\nu}}$ 
and $\vec{n_{\mu}}$
are normal vectors along their momenta   
 and  $\Phi \simeq 0.3$ is a 
kinematical factor.  T-invariance 
constrains  
the parameter  $\xi$ to a real value, i.e.,  
a non-vanishing 
Im$(\xi)$
would signal a violation of T-invariance. The value of Im$(\xi)$ 
depends on  
the model. In the case of models with non-standard scalar 
interactions, it is 
proportional to the imaginary part of the scalar coupling values, and 
measurement of $P_T$ is a    
very efficient instrument to constrain these 
models (multi-Higgs, leptoquark, minimal SUSY with R-parity 
violation). In the minimal
three Higgs Doublet Model, the 
indirect limits on $P_T$ obtained from the measurements of the 
neutron dipole moment and 
$B\rightarrow X\tau\nu$ and $b\rightarrow s\gamma$ decays are updated 
in Ref.~\cite{diwan}. 

In $K^+\rightarrow\mu^+\nu\gamma$ decay, the T-violating polarization 
is
sensitive to new pseudo-scalar, vector and axial-vector interactions. 
It can arise from the interference between the tree level amplitude 
in the SM   
 and the new CP violating amplitudes. The 
polarization can be as large as 
$10^{-2}-10^{-3}$ in models with  
 left-right symmetry, multi-Higgs 
bosons, SUSY,
and leptoquarks~\cite{geng}. As shown in~\cite{kob}, it is important 
to study both decays because
of existing correlations between T-violating  
polarizations that could 
allow one  
to 
distinguish new sources of CP violation if $P_T \neq 0$ in one or 
both decays,
or put new constraints on new CP phases if $P_T = 0$. 
\section{Detector}

\begin{figure}[htb]
\includegraphics[width=15cm,angle=0]{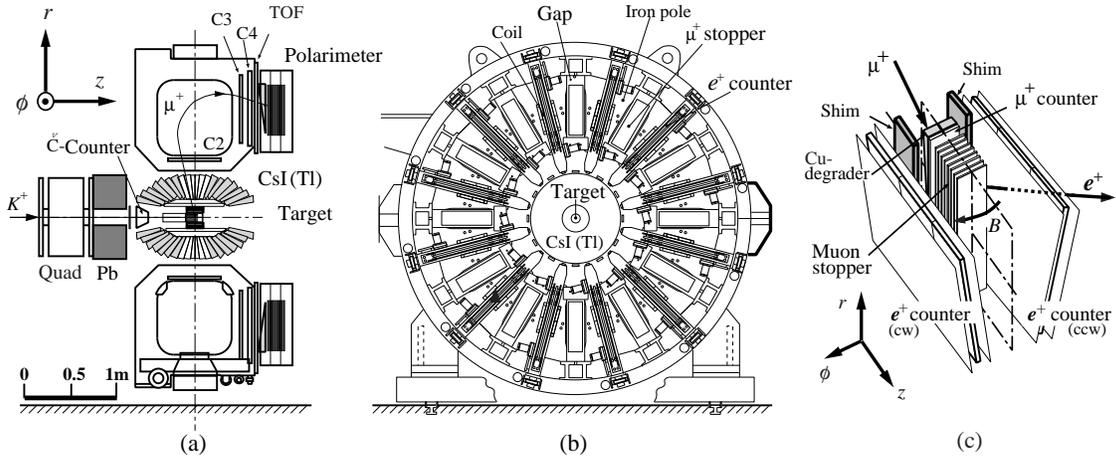} 
\caption{The layout of the KEK E246 detector:(a) side view, (b) end 
view and (c)
 one sector of the polarimeter.}
\label{e246}
\end{figure}

Fig.~\ref{e246} shows the E246 set-up. A separated $K^{+}$ beam 
($\pi/K \simeq 6$) of 660 MeV/$c$  
 is produced at the KEK 12-GeV  
 proton 
synchrotron with a typical intensity of $3.0\times 10^5$ kaons per 
0.6-s  
spill duration with a repetition of 3 s.  
A ${\rm \check{C}}$erenkov counter with a multiplicity trigger 
distinguished $K^{+}$s from 
$\pi^+$s. Kaons were then slowed in a BeO degrader, and stopped in a 
target made of 256 scintillating 
fibers located at the center of a 12-sector  superconducting toroidal 
spectrometer.
A $K^{+}_{\mu 3}$ event was identified by analyzing the $\mu^+$  
with the 
spectrometer
and detecting the $\pi^0$  
 with a CsI(Tl) photon detector consisting of 
768 modules~\cite{csi}. 
In the CsI(Tl) barrel there were twelve holes for muons to pass into 
the magnet. 
Since the solid angle coverage of the CsI(Tl) was only 75\% of 
$4\pi$, a $\pi^0$ was identified 
not only as two photons ($2\gamma$) giving a $\pi^0$ invariant mass 
($M_{\gamma\gamma}$), but also 
as one photon ($1\gamma$) with an energy greater than 70 MeV which 
preserved the directional information of 
the parent $\pi^0$. Charged particles from the target were tracked by 
means of   
 multiwire proportional chambers at the 
entrance (C2) and exit (C3 and C4) of each magnet sector,  
along with the 
target and a  
 scintillation ring hodoscope~\cite{hodoscope} around 
the target. 
The momentum resolution ($\sigma_p$=2.6 MeV/$c$  
 at 205 MeV/$c$)   
was adequate to remove the predominant background of 
$K^{+}\rightarrow\pi^{+}\pi^{0}$ ($K_{\pi2}$) decay. 
Positrons from $K^{+}\rightarrow\pi^{0} e^{+} \nu$  in the relevant 
momentum region  100-190 MeV/c were rejected by time-of-flight. 
The $K^{+}\rightarrow\pi^{+}\pi^{0}\pi^{0}$ background was 
negligible  
because the $\pi^+$ stopped in the Cu degrader and could not reach 
the polarimeter. 
 Muons entering the polarimeter (Fig.~\ref{e246}c) were degraded by a 
Cu block 
and stopped in a stack  
of pure Al plates.   
Positron counters with three layers of plastic scintillators were 
located between the magnet gaps. 
The time spectra of $e^+$ were   
 recorded by multi-stop TDCs  
 up to $20 \mu$s.   
Background associated with the beam was suppressed by a veto counter 
system surrounding the beam region. 

$P_T$ is directed in a screw sense  
around the beam axis  and 
generates 
an asymmetry $A_T=[N_{cw}-N_{ccw}]/[N_{cw}+N_{ccw}]\approx 
[N_{cw}/N_{ccw}-1]/2$ in the counting rate
between clockwise ($cw$) and counter-clockwise ($ccw$) emitted 
positrons.
Here, $N_{cw}$ and  $N_{ccw}$ are the sums of $cw$ and $ccw$ positron
counts over all 12 sectors. The sign of $P_{T}$ of
forward-going $\pi^{0}$ events is opposite to that of backward-going 
$\pi^{0}$
events. This allows one   
to take a double ratio between these two types of 
events, which is essential for reduction of most systematic errors. 
$P_{T}$ is related to $A_T$ by
$A_T = \alpha f P_{T}$, where $\alpha$ is the analyzing power of the 
polarimeter,
and $f$ is an average angular  attenuation factor.
\section{ $K^+\rightarrow\pi^0\mu^+\nu$}
\subsection{Analysis}

The experimental data are analyzed by two groups independently using
a so-called ``blind" approach.  
Each analysis selected the $K_{\mu3}$ 
events
in the $\mu^+$ momentum region of 100-190 MeV/$c$  
that removes 
the $K_{\pi2}$ decays. Most of the   
muons from pion decay in-flight in  
$K_{\pi2}$
are rejected by using the $\chi^2$ cut in tracking.
For $2\gamma$ events, the $\pi^0$ is identified by $\gamma-\gamma$ 
coincidence in the   
CsI and applying a   
 cut on the pion invariant mass. 
The $K_{e3}$  events which also satisfy these requirements  are 
removed by time-of-flight. The $K_{\mu3}$ events from  in-flight kaon 
decays in the target were rejected by applying a   
 cut on the 
kaon decay time: the   
time difference between kaon stop and decay should 
be more than 2 ns.  
The ``good" $K_{\mu3}$ events were separated 
into two classes:
$fwd$ events when the angle between $\pi^0$ and beam directions  
(z-axis) was   
less than $70^{\circ}$ and $bwd$ events when   
 the angle between $\pi^0$ 
and beam directions was   
 more than $110^{\circ}$.

The signal was extracted by integrating the positron time spectrum 
from
$\mu^+\rightarrow e^+\nu\bar{\nu}$ decays of muons stopped in 
polarimeter 
after subtraction of the background.
A null asymmetry check ($A_0=[(N_{cw}/N_{ccw})_{fwd+bwd}-1]/2 $ for 
$fwd + bwd$ events) performed in each analysis  did not show 
significant spurious asymmetry. Then the   
T-violating asymmetry $A_T$ 
\begin{equation}
A_T =\frac{1}{4} \left[ \frac{(N_{cw} / N_{ccw})_{fwd}} {(N_{cw} / 
N_{ccw})_{bwd}} - 1 \right]
\end{equation}
was obtained using a combination of both   
 analyses as described
 in Ref.~\cite{prl}.

\begin{figure}[htb]
\centering\includegraphics[width=12cm, height=12cm, angle=0]
{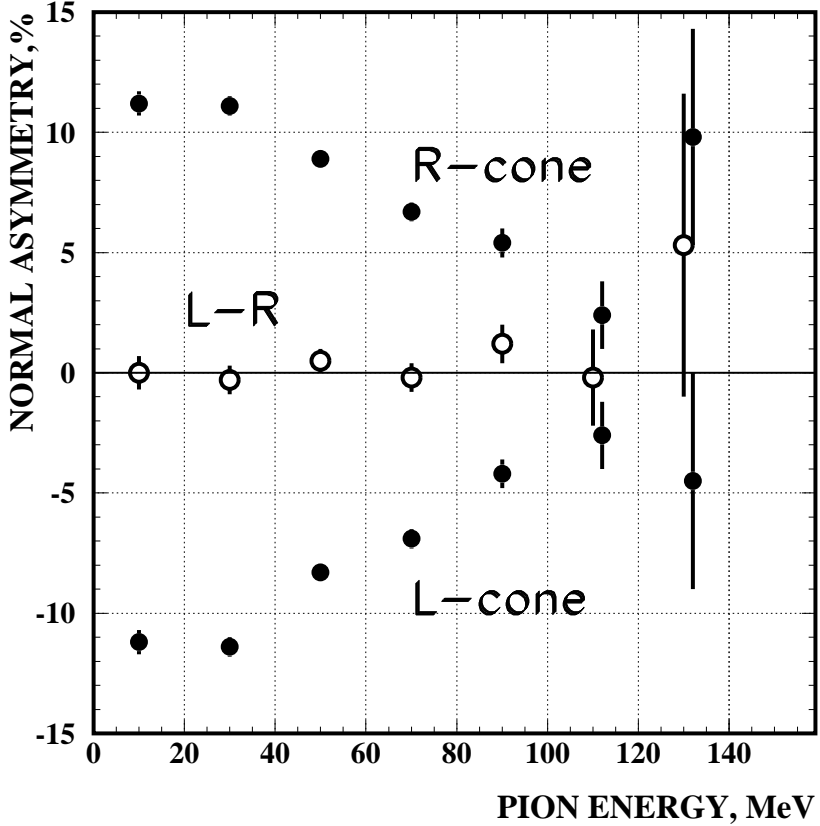}
\caption{Normal asymmetry measured for two-photon $K_{\mu3}$ events.
 {\em R-cone}
shows the events with  pion moving into the right hemisphere relative 
to median   
 plane of the given magnet sector,
{\em L-cone} are the events with  pion moving into the left 
hemisphere. Open
circles show the difference between the {\em R} and {\em L} normal 
asymmetries.}
\label{normpol}
\end{figure}

The detector sensitivity to muon polarization (analyzing power 
$\alpha$) can be 
obtained from the measurement of the normal  muon polarization, $P_N$,
which is an in-plane component of the muon polarization normal to 
muon momentum.
$P_N$ is a T-even observable ${\bf s}_{\mu}\cdot
({\bf p}_{\mu}\times({\bf p}_{\pi}\times{\bf p}_{\mu}))$. The value 
of $P_N$
averaged over the   
accepted part of the $K_{\mu3}$ Dalitz plot is about 
0.6 and can be measured if the accepted events are separated into two 
classes: events when the   
pion
moves into the left hemisphere with   
 respect to the median  
  plane of the 
given magnet sector 
 and events when the    
 pion moves into the right
hemisphere.   
 The values of $P_N$ for these two classes should be the 
same but
opposite in sign.  
In addition, $P_N$ decreases as the pion energy 
increases.  The results are presented in Fig.~\ref{normpol}.
From this measurement the value of $\alpha \simeq 0.2$ is extracted.
The difference between the left-going and right-going  
 asymmetries is consistent with zero, and   
that is also a   
good test of the detector  
azimuthal symmetry.

The contamination of the beam accidental backgrounds in ``good"  $K_{\mu3}$ events was about 8\% for $2\gamma$ events and
$\sim 9$\% for $1\gamma$ events.  The constant background in the polarimeter 
was $11-12$\%. Both these backgrounds only delute the sensitivity to $P_T$
by 10\%, but they do not produce any spurious T-violating asymmetry.
The main systematics contributions to $P_T$ come from the presence of 
two
large components  
of the  muon in-plane polarization, $P_L$ which is 
parallel to
the muon momentum and $P_N$ ($P_T\ll P_{N,L}\leq 1$).  Most of their
contribution is cancelled by the azimuthal symmetry of the detector 
as well as by  
the $fwd/bwd$ ratio. The largest  systematic errors are due to   
 the 
misalignment of
the polarimeter, the asymmetry of magnetic field distribution, and the
asymmetrical kaon stopping distribution. The $fwd/bwd$ ratio 
dramatically  
reduces these contributions. The total systematic error of $P_T$ is 
 found to be $0.9\times 10^{-3}$~\cite{prl},  
 and is  
 significantly lower than
 the statistical error.  
\subsection{Results}

\begin{figure}[htb]
{\centering\includegraphics[width=13cm, angle=0]{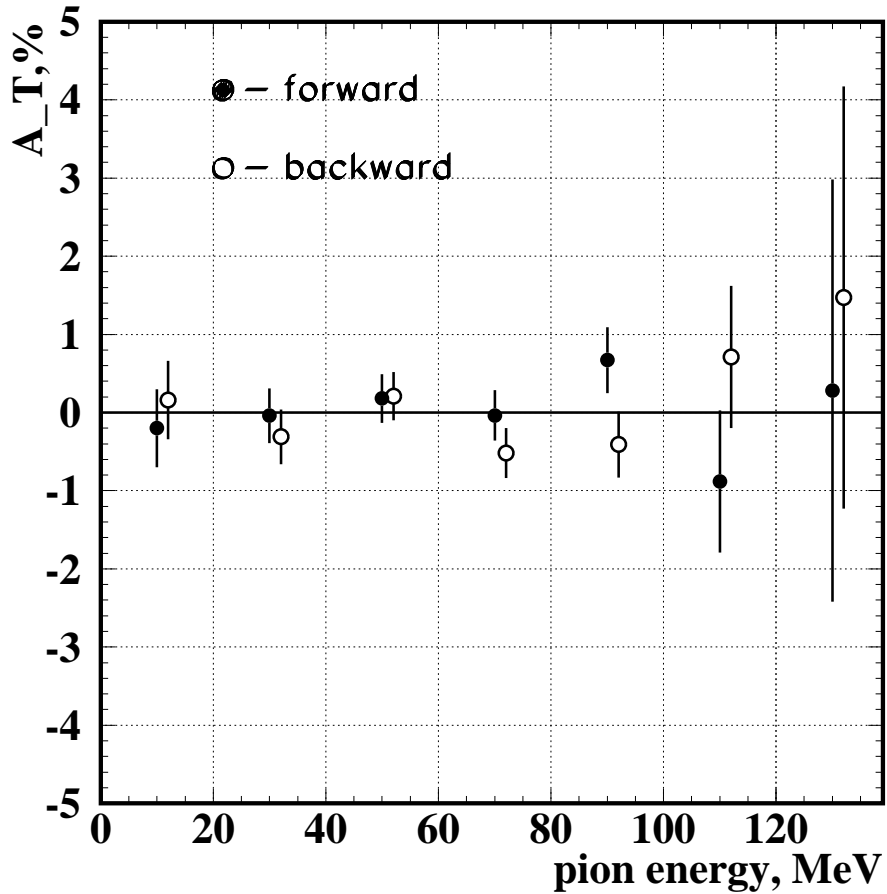}}
\caption{T-odd muon asymmetry as a function of the pion kinetic
energy for forward- and backward-going $\pi^0$s obtained by one 
analysis of  
1998 data.}
\label{pt}
\end{figure}

After the analysis of data taken in 1996 and 1997 we selected
about  $3.9 \times 10^6$ good $fwd+bwd$ $K_{\mu3}$ events. 
The value  obtained for $P_T$ was  
$-0.0042\pm 0.0049 (stat)\pm0.0009 (syst)$ 
and
Im$(\xi)= -0.013\pm 0.016 (stat) \pm 0.003 (syst)$~\cite{prl}.
Analysis of the 1998 data was recently completed.  
The results obtained in both 
analyses are consistent and both show  zero transverse muon 
polarization~\cite{aoki}. The dependence of $P_T$ as a function of
pion energy for forward- and backward-going pions ($2\gamma$ events) 
is shown in Fig.~\ref{pt}.
The statistical error of $P_T$ of the combined 1996-98 result is 
expected to be about $3\times 10^{-3}$ (1$\sigma$ level) that 
corresponds to $\delta$Im$(\xi) \sim 1.1\times 10^{-2}$.
The systematic error is three times  smaller than the statistical 
one.  
\section{$K^+\rightarrow\mu^+\nu\gamma$}
\subsection{Analysis}

\begin{figure}[htb]
\centering\includegraphics[width=12cm,angle=0]{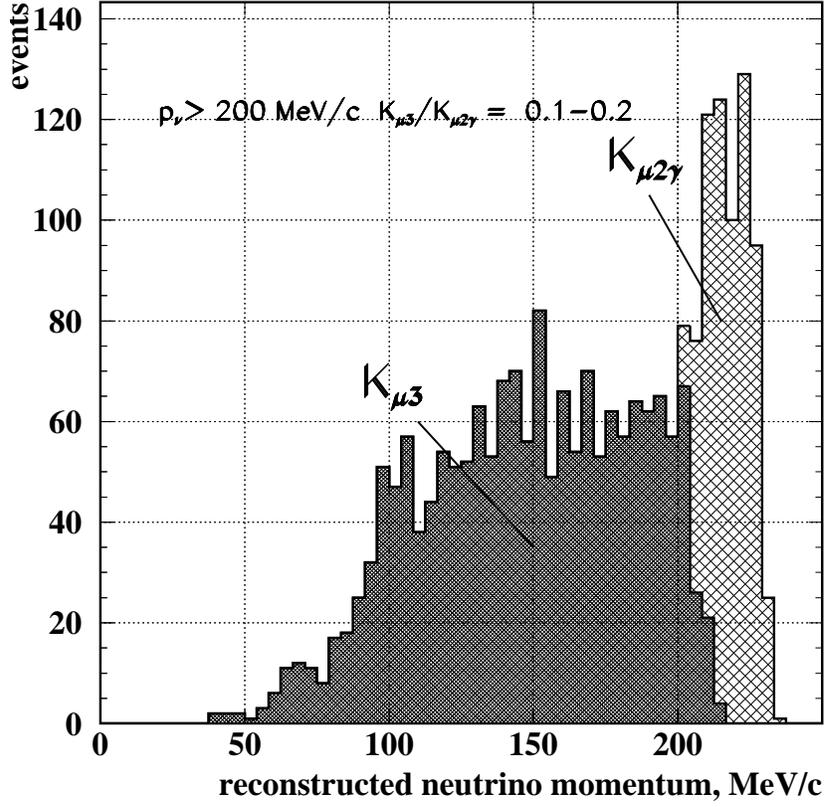} 
\caption{Momentum of the missing particle of 1$\gamma$ events after 
applying cuts 
$ |M^2_{miss}|\leq 5000 $ MeV$^2$/c$^4$, ${\bf p}_{\mu}\leq$195 
MeV/c  and  $E_{\gamma}> 50$ MeV. The distribution with dominating
contribution from $K_{\mu2\gamma}$ decay is obtained after the 
additional cut $\Theta_{\mu\gamma} \leq 75^{\circ}$.  $K_{\mu3}$ 
events 
are selected applying the criteria $\Theta_{\mu\gamma} \geq 
75^{\circ}$ }
\label{kmu2g}
\end{figure}

\begin{figure}[htb]
\centering\includegraphics[width=12cm,angle=0]{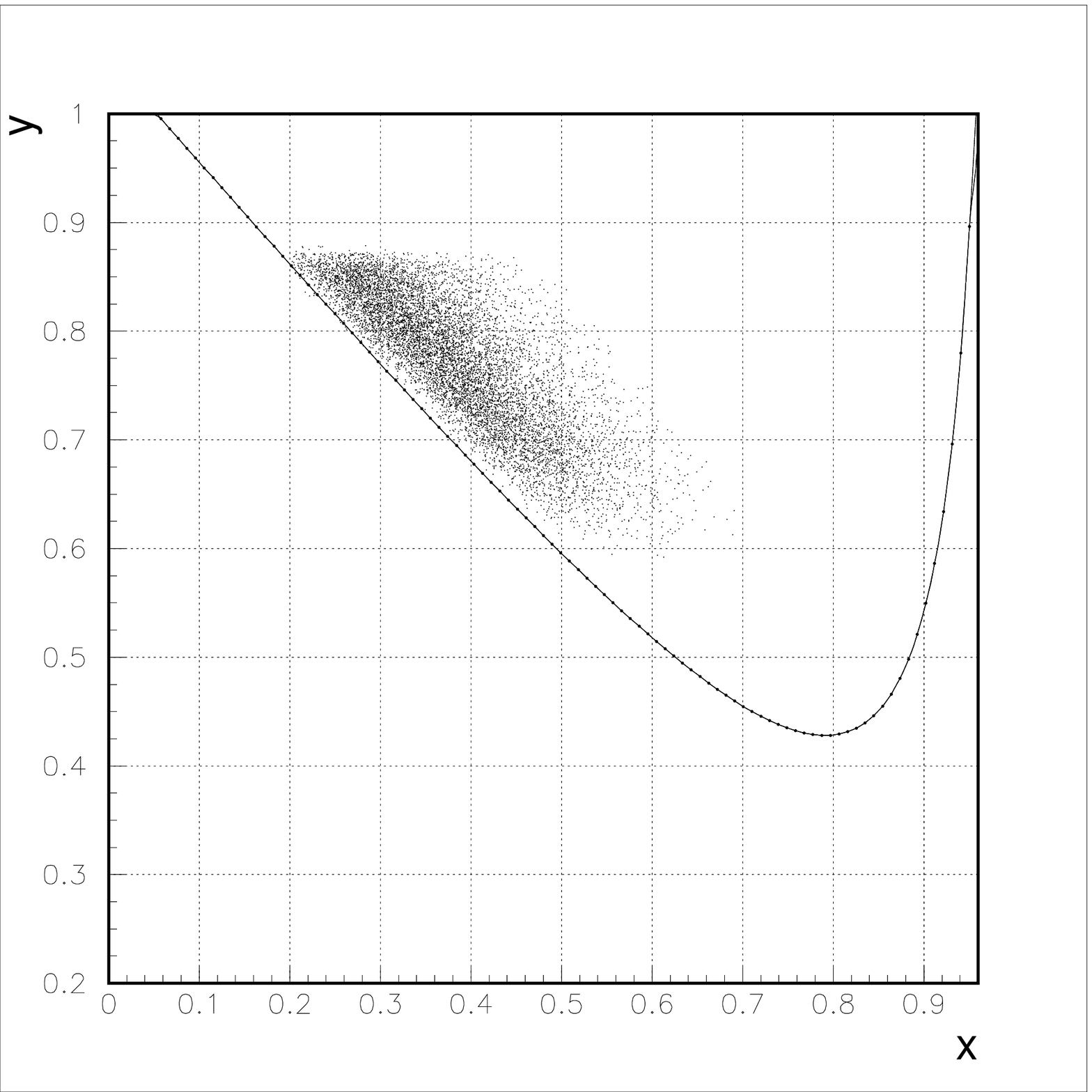} 
\caption{ Distribution of accepted $K_{\mu2\gamma}$ events. Here,
$x = 2E_{\gamma}/m_K$ and $y = 2E_{\mu}/m_K$ are normalized to kaon mass
energies of photon and muon, respectively. The solid curve encloses
the area of the $K_{\mu2\gamma}$ Dalitz plot. }
\label{dalitz}
\end{figure}

The key problem in the extraction of  
$K^+\rightarrow\mu^+\nu\gamma$ events
is the presence of the intense background from
the $K_{\mu3}$ decays when only one photon from the $\pi^0$ decay 
was  
detected and the second  
 disappeared in one  of 12  holes in the 
CsI calorimeter.
However, 
due to the   
 precise measurement of the muon momentum and the photon 
energy and direction, the kinematics  
of  
 $K_{\mu3}$ and  
$K_{\mu2\gamma}$ decays can be reconstructed completely.   The 
kinematical parameters such as missing mass squared
$M^2_{miss}$, the angle between the muon and photon,   
$\Theta_{\mu\gamma}$,   
 and
the neutrino momentum, $p_{\nu}$,   
 can be efficiently used  
to suppress   
the $K_{\mu3}$ background. The missing mass squared 
for 1$\gamma$ events   
is
\beqn
\label{missmass}
M^2_{miss} = E^2_{miss} - {\bf p}^2_{miss} = (m_K - E_{\mu} 
-E_{\gamma})^2 -
({\bf p}_K - {\bf p}_{\mu} - {\bf q})^2.
\eeqn
For  $K_{\mu2\gamma}$ events, the neutrino is the only missing 
particle, therefore, ${\bf p}_{\nu} = {\bf p}_{miss}$, $E_{\nu} = 
E_{miss}$ and 
$M^2_{miss} = E_{\nu}^2 - {\bf p}^2_{\nu} = 0$, while the 
$M^2_{miss}$ of  
 $K_{\mu3}$ events is   
 distributed over a   
  wide range 
with the maximum of the broad 
peak at
about 20000  MeV$^2$/$c^4$.   
 The best suppression of  
$K_{\mu3}$ events is reached for a cut on $ |M^2_{miss}|\leq 
5000 $  
MeV$^2$/$c^4$.   
 Additional  
 to the $M^2_{miss}$ cut is the cut on  
  ${\bf p}_{\nu} = -({\bf p}_{\mu} + {\bf 
p}_{\gamma})$.   
Other  
cuts are the  muon momentum, ${\bf p}_{\mu}\leq$195 MeV/$c$  
and 
$\Theta_{\mu\gamma} \leq 75^{\circ}$. 
The reconstructed neutrino momentum of $1\gamma$ events after 
imposing the 
 $M^2_{miss}, \Theta_{\mu\gamma},{\bf p}_{\mu}$ cuts and 
 accepting events with $E_{\gamma}> 50$ MeV is given in 
Fig.~\ref{kmu2g}.  
The peak at $p_{\nu}\sim 220$ MeV/$c$    
corresponds to $K_{\mu2\gamma}$ 
events.   In the region with $p_{\nu}>200$ MeV/$c$,  
 the contamination  of the $K_{\mu3}$ events is estimated to be 
$\leq$20\%. The accepted $K_{\mu2\gamma}$
 events are concentrated in the Dalitz plot region where the 
 inner bremsstrahlung (IB)  
 term 
 dominates and
 the normal polarization of the muon in the decay plane, $P_N$,   
  is about 0.2, as shown in Fig.~\ref{dalitz}. 
  
 It is interesting that $P_N$ in $K_{\mu2\gamma}$ and $K_{\mu3}$ have 
the
 opposite signs. This provides another test of the  $K_{\mu2\gamma}$ 
events
 and shows the detector sensitivity to
 polarization measurements in $K_{\mu2\gamma}$ decay. For the data  
accumulated in 1998
the values obtained for  
normal asymmetries for $1\gamma$ $K_{\mu3}$ 
decays for left and right cones were   
$A^L_N = (-4.89 \pm 0.2) \times 
10^{-2}$ and  
$A^R_N = (4.36 \pm 0.2) \times 10^{-2}$.  
The 
same values  for $K_{\mu2\gamma}$ events
 are $A^L_N = (0.64 \pm 1.16) \times 10^{-2}$ and $ A^R_N = (-2.05 
\pm 1.16) \times 10^{-2}$. All these quantities are in agreement with 
expected 
polarization values, 
 although the statistical errors are too   
  large in the 
$K_{\mu2\gamma}$ decay to support a   
 more definite conclusion.
\subsection{Results}

 From the  data accumulated by E246 in 1996-2000, 
 about $(2-3)\times 10^5$ ``good" $K_{\mu2\gamma}$ events are 
estimated to be 
 extracted.
The final sensitivity to T-odd muon polarization in $K_{\mu2\gamma}$ 
decay is expected to be   
at the
level of about $1.5\times 10^{-2}$. The analysis is in   
 progress.
 \section{Future Prospects}   

The final E246 sensitivity to $P_T$ in $K_{\mu3}$ is expected to be  
 $\sim 2\times 10^{-3}$ 
 corresponding   
  to $\delta$Im$(\xi)$ of about
$7\times 10^{-3}$. Since statistics mainly determines the sensitivity,
the detector still has   
 a capability to  improve the sensitivity 
at  least by a factor of 2 using a   
more intense kaon beam.

New planned experiments will reach higher sensitivity in
measuring both $K_{\mu3}$ and $K_{\mu2\gamma}$ decays.
A  proposed E923 experiment at BNL~\cite{bnl} is designed to use 
in--flight 
$K^+$ decays.  A cylindrical active polarimeter
around the kaon beam and an electromagnetic calorimeter will be used  
to
reconstruct $K_{\mu3}$ decays and suppress background.
The detector acceptance to $K_{\mu3}$ events is about 
2.5$\times10^{-5}$ per  2-GeV/$c$    
 kaon.
The advantage of the in--flight experiment is thus relatively 
high detector acceptance. The statistical sensitivity (1$\sigma$ 
level) 
to $P_T$ in this experiment will be about 1.3$\times10^{-4}$ which 
corresponds 
to  $\delta$Im$(\xi)=7\times10^{-4}$. In this experiment sensitivity 
of 
$\leq10^{-3}$ can be also obtained for $P_T$ in $K_{\mu2\gamma}$ 
decay.

A new approach to measure the  T-odd polarization  in $K_{\mu3}$  
and  $K_{\mu2\gamma}$   with a statistical sensitivity to
 $P_T$ at the 1$\sigma$ level of 
 about $10^{-4}$ using stopped kaons has recently been 
proposed~\cite{kudenko}.
The virtue of this   experiment is a high 
resolution measurement of the $\pi^0$ (momentum resolution of $\sim$ 
1--2\%) 
 that will allow almost complete suppression of   
  $K_{\pi2}$ decay. 
 Another important detector element is an active muon polarimeter 
which  provides higher  sensitivity to the muon polarization and 
more efficient   
 background suppression than a   
 passive polarimeter by means of   
 complete measurement of the muon track,  
 muon stopping point,   
  and both 
energy and direction of the   
positron from muon decay.  
A calorimeter  and  an additional  highly efficient photon veto 
system around the    
polarimeter cover nearly  4$\pi$ solid angle. This experiment can be 
done at the low energy separated kaon
beam at JHF~\cite{jhf}.
\section{Conclusion} 

The status of measurements of T-violating  
muon polarization in the 
decays $K^+\rightarrow\pi^0\mu^+\nu$ and 
$K^+\rightarrow\mu^+\nu\gamma$ and
future prospects    
 are outlined.
These measurements with sensitivity to $P_T$ at the level of 
$10^{-3}-10^{-4}$ could be a good test of non-standard sources of CP 
violation.

This work was supported by the
Russian Foundation for Basic Research Grant No.~99--02--17814.

\end{document}